\documentclass[useAMS,usenatbib]{mn2e}
\usepackage{graphicx}
\usepackage{aas_macros}

\newcommand{\kms}{\mbox{${\rm km\,s}^{-1}$}}

\newcommand{\Msolar}{\mbox{${M}_{\sun}$}}
\newcommand{\Lsolar}{\mbox{${L}_{\sun}$}}
\newcommand{\Rsolar}{\mbox{${R}_{\sun}$}}

\newcommand{\J}{\mbox{J0247$-$25}}
\newcommand{\logg}{\mbox{$\log\,{\rm g}$}}
\newcommand{\Teff}{\mbox{T$_{\rm eff}$}}

\title[J0247$-$25]{Discovery of a stripped red giant core in a bright
eclipsing binary system\thanks{Based on observations made with ESO
Telescopes at the La Silla Observatory under programme ID
084.D-0348(A).}}
\author[P.F.L. Maxted et~al.]{P.F.L.~Maxted$^{1}$\thanks{E-mail:
pflm@astro.keele.ac.uk}, D.R.~Anderson$^{1}$, M.R.~Burleigh$^2$, A. Collier
Cameron$^3$,\newauthor    U.~Heber$^{4}$,
B.T.~G\"{a}nsicke$^5$,
S.~Geier$^4$, T. Kupfer$^{4}$, T.R.~Marsh$^{5}$, G.~Nelemans$^6$,\newauthor   
S.J.~O'Toole$^7$,
R.H.~{\O}stensen$^8$, B.~Smalley$^{1}$, R.G. West$^{9}$\\
$^1$Astrophysics Group,  Keele University, Keele, 
      Staffordshire ST5 5BG\\
$^2$Department of Physics and Astronomy, University of Leicester, University
Road, Leicester LE1 7RH\\
$^3$SUPA, School of Physics and Astronomy, University of St.\
Andrews, North Haugh,  Fife, KY16 9SS, UK\\
$^4$Dr. Karl Remeis-Observatory \& ECAP, Astronomical Institute,
Friedrich-Alexander University Erlangen-Nuremberg,\\ Sternwartstr.~7,
D~96049 Bamberg, Germany\\
$^5$Department of Physics, University of Warwick, Coventry, CV4 7AL \\
$^6$Department of Astrophysics, IMAPP, Radboud University Nijmegen, PO Box
9010, 6500 GL Nijmegen, The Netherlands \\
$^7$Australian Astronomical Observatory, PO Box 296, Epping, NSW, 1710,
Australia\\
$^8$Institute of Astronomy, K.U.Leuven, Celestijnenlaan 200D, 3001, Heverlee,
Belgium\\
$^9$Department of Physics and Astronomy, University of Leicester, University
Road, Leicester LE1 7RH
}

\date{To be inserted}
\pagerange{\pageref{firstpage}--\pageref{lastpage}} \pubyear{2009}

\begin{document}

\maketitle

\label{firstpage}

\begin{abstract}
 We have identified a star in the WASP archive photometry with an unusual
lightcurve due to the total eclipse of a small, hot star by an apparently
normal A-type star and with an orbital period of only 0.668\,d. From an
analysis of the WASP lightcurve together with V-band and I$_{\rm C}$-band
photometry of the eclipse and a spectroscopic orbit for the A-type star we
estimate that the companion star has a mass of $0.23\pm0.03$\Msolar\ and a
radius of $0.33\pm0.01$\Rsolar, assuming that the A-type star is a
main-sequence star with the metalicity appropriate for a thick-disk star. The
effective temperature of the companion is $13\,400\pm 1\,200$K from which we
infer a luminosity of $3\pm1$\Lsolar. From a comparison of these parameters to
various models we conclude that the companion is most likely to be the remnant
of a red giant star that has been very recently stripped of its outer layers
by mass transfer onto the A-type star. In this scenario, the companion is
currently in a shell hydrogen-burning phase of its evolution, evolving at
nearly constant luminosity to hotter  effective temperatures prior to ceasing
hydrogen burning and fading to become a low-mass white dwarf composed of
helium (He-WD). The system will then resemble the pre-He-WD/He-WD companions
to A-type and B-type stars recently identified from their Kepler satellite
lightcurves (KOI-74, KOI-81 and KIC~10657664). This newly discovered binary
offers the opportunity to study the evolution of a stripped red giant star
through the pre-He-WD stage in great detail. 

\end{abstract}


\section{Introduction}

 Wide-area surveys for transiting extra-solar planets such as WASP (Wide Angle
Search for Planets, \citealt{2006PASP..118.1407P}),  HATnet
\citep{2004PASP..116..266B}, XO \citep{2005PASP..117..783M} and TrES
\citep{2006AAS...20922602O} provide high cadence photometry for millions of
bright stars across a large fraction of the sky. This provides the opportunity
to find and study many new examples of known classes of variable star, e.g.,
eclipsing brown dwarf binary systems \citep{ 2011ApJ...726L..19A}, double-mode
RR~Lyr stars \citep{2010IBVS.5955....1W}, W~UMa stars
\citep{2011A&A...528A..90N}, young solar-type stars
\citep{2011arXiv1104.2986M} and cataclysmic variable stars
\citep{2011JAVSO.tmp..140W}. New discoveries will certainly be made now that
much of the data from these surveys is becoming widely available
\citep{2010A&A...520L..10B}. The photometric precision achieved by these
surveys with modest equipment ($\la 0.01$\,magnitudes at V$\approx$12) is
impressive, but cannot compete with the micro-magnitude photometry acheived
from space by surveys such as CoRoT \citep{2006ESASP1306...33B} and Kepler
\citep{2009IAUS..253..289B}. Photometry with this precision has made  it
possible to identify new types of variable star that are difficult or
impossible to study from the ground, e.g., triply eclipsing binary stars
\citep{ 2011Sci...331..562C}, stars with tidally excited pulsations \citep{
2011arXiv1102.1730W}, subdwarfs with white dwarf companions
\citep{2011MNRAS.410.1787B},  and white dwarf companions to early-type main
sequence stars \citep{2010ApJ...713L.150R, 2010ApJ...715...51V}.

 One advantage that ground-based surveys currently have over space-based
surveys is that they cover a much larger fraction of the sky. This makes is
possible in some cases to discover rare or extreme examples of these new
classes of variable star. In the case of the white-dwarf companions to
early-type stars KOI-74 and KOI-81, these were identified from the eclipses and
transits in the Kepler lightcurves, even though these features are  much less
than 1\% deep \citep{2010ApJ...713L.150R, 2010ApJ...715...51V}. These low-mass
white dwarfs have an unsual evolutionary history, but it is difficult to study
them in detail because they are very much fainter than their companions.
Ground-based surveys offer the opportunity to discover similar eclipsing
binary systems that are more favourable for detailed follow-up observations.

 Low mass white dwarf stars ($M<0.4\Msolar$) are the product of binary
star evolution \citep{1993PASP..105.1373I, 1995MNRAS.275..828M}. They are the
result of mass transfer from a red giant onto a companion star when the giant
has a small degenerate helium core. There are several possible outcomes  from
this mass transfer depending on the mass ratio of the binary and the type of
companion star. If the companion is a neutron star then the mass transfer is
likely to stable so the binary can go on to become a low mass X-ray binary
(LMXB) containing a millisecond pulsar. Several millisecond radio pulsars are
observed to have low mass white dwarf (LMWD) companions
\citep{2008LRR....11....8L}. Many new LMWDs have recently been identified in
the Sloan Digital Sky Survey \citep{2007ApJ...660.1451K} and from proper
motion surveys \citep{2009A&A...506L..25K}.  Searches for radio pulsar
companions to these LMWDs have so far found nothing, suggesting that the
majority of these LMWDs have white dwarf companions
\citep{2009ApJ...697..283A}. LMWD can also be produced by mass transfer from a
red giant onto a main sequence star, either rapidly through unstable
common-envelope envolution or after a longer-lived ``Algol'' phase of stable
mass transfer \citep{1969A&A.....1..167R, 1970A&A.....6..309G,
2004A&A...419.1057W, 1993PASP..105.1373I, 2003MNRAS.341..662C,
2001ApJ...552..664N}.

 The evolution of LMWDs is expected to be very different from more massive
white dwarfs.  It is expected that once mass transfer stops the LMWD will have
a thick layer of hydrogen surrounding the degenerate helium core which leads
to steady hydrogen shell burning via the p-p chain and can lead to unstable
phases of CNO burning for LMWD in the mass range 0.2\,--\,0.3\Msolar\  
\citep{1999A&A...350...89D}.  These hydrogen shell flashes
lead to mixing between the inner and outer layers, producing a hydrogen
deficient surface composition for the LMWD. Models that include hydrogen shell
burning provide a much better match between the ``cooling age'' of the LMWD
and the ``spin-down age'' of the millisecond pulsars in LMXBs. The number of
hydrogen shell flashes depends strongly on the mass of the hydrogen layer that
remains on the surface of the LMWD \citep{2000MNRAS.316...84S}, which
in-turn depends strongly on the details of the mass loss from the red giant
\citep{2002ApJ...565.1107P}. Details such as the treatment of
diffusion can also lead to large differences in the predictions between
different models \citep{2001MNRAS.323..471A}. 

 In this paper we present the discovery from WASP photometry of an eclipsing
binary star that is related to KOI-74, KOI-81 but that has
much deeper eclipses, i.e. the companion to the early-type star is much
brighter and larger than the white-dwarf companions to these Kepler
discoveries. We present the data used to identify this new eclipsing binary
star and the follow-up photometry and spectroscopy we have obtained; we
analyse these data to determine the masses, radii and luminosities of the
stars; and we outline how our discovery of this bright pre-white dwarf
companion to an early-type star makes it possible to study the formation of an
LMWD in detail.

\begin{table}
 \caption{Catalogue photometry and astrometry of \J. The GALEX FUV and NUV
fluxes are given as AB magnitudes f$_{\rm AB}$ and n$_{\rm AB}$, respectively.
Sources are GALEX \citep{2007ApJS..173..682M}, NOMAD \citep{
2004AAS...205.4815Z} PPMXL\citep{2010AJ....139.2440R}, UCAC3 \citep{
2010AJ....139.2184Z},  2MASS \citep{2006AJ....131.1163S}.}
 \label{mags}
 \begin{tabular}{@{}lrl}
  \hline
  Parameter &\multicolumn{1}{l}{Value} & Source \\
  RA (J2000.0) & 02 47 43.38 & NOMAD \\
  Dec (J2000.0) &  $-$25 15 49.3 & NOMAD \\
  $\mu_{\alpha}$ & $23.2 \pm 2.5$ mas/yr & PPMXL \\ 
                 & $25.9 \pm 1.2$ mas/yr & UCAC3 \\ 
  $\mu_{\delta}$ & $-5.4 \pm 1.2$ mas/yr & PPMXL \\ 
                 & $-5.3 \pm 1.0$ mas/yr & UCAC3 \\ 
  f$_{\rm AB}$  & $16.03 \pm 0.02$ &  GALEX \\
  n$_{\rm AB}$  & $14.46 \pm 0.01$ & GALEX \\
  B & 12.13 & NOMAD \\
  V & 12.44 & NOMAD \\
  J & 11.85 & 2MASS \\
  H & 11.81 & 2MASS \\
  K$_s$ & 11.79 & 2MASS \\
\hline
\end{tabular}
\end{table}

\begin{figure} 
\includegraphics[width=0.45\textwidth]{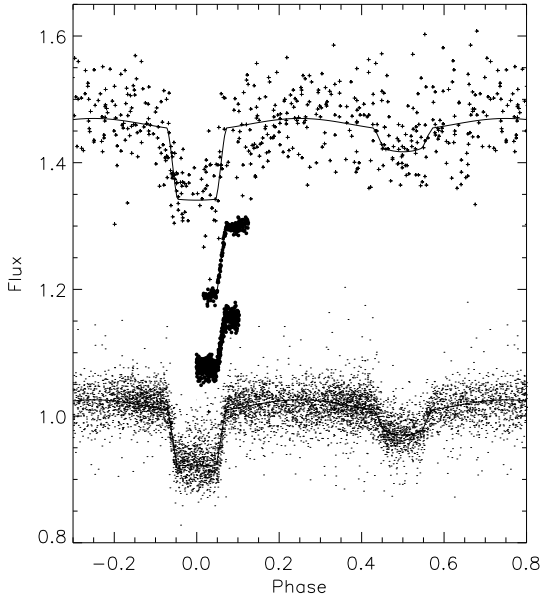} 
\caption{Lightcurves of \J\ with lightcurve model fits (solid lines). From
bottom-to-top: WASP (small points); SAAO 1.0-m I$_{\rm C}$-band and V-band
(filled circles); ASAS V-band (small crosses). 
\label{lcfit}  } 
\end{figure} 

\begin{figure} 
\includegraphics[width=0.45\textwidth]{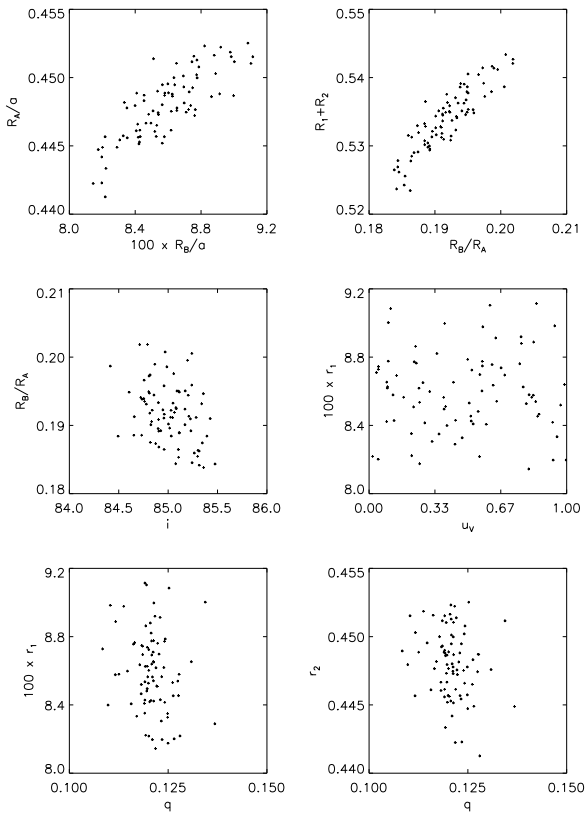} 
\caption{Distribution of selected lightcurve parameters from the bootstrap 
Monte Carlo simulation.
\label{plotpar6}  } 
\end{figure} 

\begin{figure} 
\includegraphics[angle=270,width=0.45\textwidth]{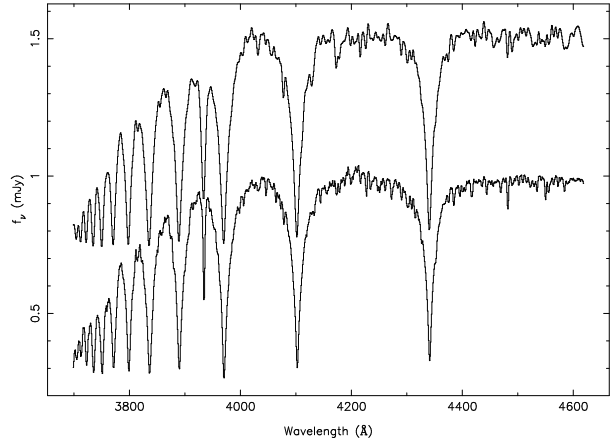} 
\caption{Flux calibrated and normalized GMOS-S spectrum of \J\ (lower
spectrum) compared to the
normalized, flux-calibrated and smoothed spectrum of the A6Vp star HD148898
offset by +0.5 units (upper spectrum).
\label{spec}  } 
\end{figure}

\begin{figure} 
\includegraphics[width=0.45\textwidth]{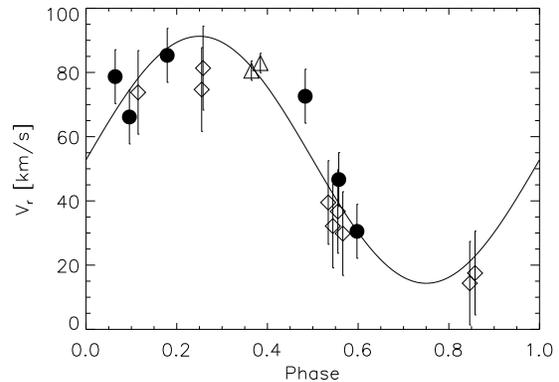} 
\caption{Radial velocity measurements of J0247$-$25\,B as a function of phase 
with a circular orbit fit (solid line). The spectrograph used is indicated as
follows: filled circles -- EFOSC2; triangles -- ISIS; diamonds -- GMOS.
\label{rvfit}  } 
\end{figure}

\section{Observations and data reduction}

\subsection{WASP photometry}
 The star, 1SWASP~J024743.37$-$251549.2 (\J\ hereafter) was observed
by the WASP-South instrument as part of the WASP survey. The WASP survey is
described in \citet{2006PASP..118.1407P} and \citet{2008ApJ...675L.113W}. The
data from this survey are automatically processed and analysed in order to
identify stars with lightcurves that contain transit-like features that may
indicate the presence of a planetary companion.  The  candidate selection
methods can be found in \citet{2007MNRAS.380.1230C},
\citet{2008MNRAS.385.1576P}, and references therein. In practice, these
automatic methods produce tens of thousands of candidates, so we use a
database to store the results of the automatic analysis plus other information
available for the stars such as catalogue photometry and astrometry. This
makes it possible to efficiently reject large numbers of  candidates that are
unlikely to host planets using a variety of criteria such as eclipse depth and
the noise level in the lightcurve. The number of candidates that remain after
sifting is small enough to make selection by inspection of the available data
by a small number of people feasible. The same database can also be used to
identify eclipsing binary stars by modifying the sifting criteria.

\J\ was spotted by one of us (PM) while looking at lightcurves of stars
with deep eclipses and low reduced proper motions, i.e., stars that may be
eclipsing  binary subdwarfs. Catalogue photometry and astrometry of \J\
are summarized in Table~\ref{mags}. The automatic transit detection algorithm
correctly identified a period of 0.6678\,d from 6633 observations of this star
obtained with the WASP-South instrument. The observations were obtained with a
single camera  through a broad-band filter (400\,--\,700\,nm) between 2006
August 10 and 2007 December 31. The WASP photometry is shown as a function of
orbital phase in Fig.~\ref{lcfit}. The deeper of the two eclipses in the
lightcurve shows a flat section between a sharp ingress and egress. This type
of lightcurve is produced by the eclipse of one star by a larger but cooler
star. It is not possible to produce a lightcurve with these properties if both
the stars in the binary are on the main sequence.  For this reason we
organised follow-up observations of this unusual object.

\subsection{SAAO 1.0-m photometry}
 We observed the egress phases of two eclipses of \J\ using the UCT CCD
photometer on the SAAO 1.0-m telescope. The star approximately 2.5 magnitudes
fainter located 71\,arcsec west of \J\ was used as a comparison star. Images
with an exposure time of 10\,s through an I$_{\rm C}$-band filter were
obtained on the night 2009 October 30.  Exposures with an exposure time of
30\,s  through a V-band filter were obtained on the night 2009 November 5. We
used synthetic aperture photometry to measure the apparent fluxes of \J\ and
the comparison star. The apparent flux of \J\ relative to the comparison star
normalized to the flux out of eclipse is shown in Fig.~\ref{lcfit}. 
\begin{table}
 \caption{Parameters for the lightcurve model fit by least-squares. Parameter
definitions are given in the text.}
\label{lcfitTable}
 \begin{tabular}{@{}lr}
\hline
  Parameter &\multicolumn{1}{l}{Value}   \\
\hline
${\rm T}_0 $&$    2455135.29504\pm   0.00004 $\\
P [d]&$    0.6678321 \pm   0.0000002     $ \\
$S_{\rm ASAS} $&$      3.36 \pm     0.16 $ \\
$S_{\rm WASP} $&$      2.50\pm     0.14$ \\
$S_I $&$      2.16 \pm     0.12 $ \\
$S_V $&$      3.17 \pm     0.15 $ \\
$b $&$     0.163 \pm   0.007 $ \\
$R_{\rm A}/a $&$     0.4492 \pm   0.0025 $ \\
$R_{\rm B}/a $&$    0.0866 \pm   0.0023 $ \\
$k $&$     0.1929 \pm   0.0043 $ \\
$q $&$     0.121  \pm   0.005  $ \\
$i$ [$^{\circ}$]&$      85.0 \pm     0.2  $ \\
\hline
\end{tabular}
\end{table}

\begin{table}
 \caption{Heliocentric radial velocities for \J\,A.}
 \label{rvtable}
 \begin{tabular}{@{}rrl}
  \hline
  \multicolumn{1}{l}{HJD} & \multicolumn{1}{l}{$V_r$} 
& \multicolumn{1}{l}{Instrument}\\
  \multicolumn{1}{l}{$-$2450000} & \multicolumn{1}{l}{[\kms]} & \\
\hline
 5137.5367 &$ 80.5 \pm  3.0$& ISIS \\
 5137.5499 &$ 82.9 \pm  3.0$& ISIS \\
 5144.6820 &$ 78.7 \pm  8.4$& EFOSC2 \\
 5144.7587 &$ 85.3 \pm  8.4$& EFOSC2 \\
 5145.6794 &$ 46.6 \pm  8.4$& EFOSC2 \\
 5145.7060 &$ 30.5 \pm  8.4$& EFOSC2 \\
 5146.7065 &$ 66.1 \pm  8.4$& EFOSC2 \\
 5147.6335 &$ 72.5 \pm  8.4$& EFOSC2 \\
 5128.6879 &$ 74   \pm 13$& GMOS-S \\
 5129.8440 &$ 14   \pm 13$& GMOS-S \\
 5129.8522 &$ 18   \pm 13$& GMOS-S \\
 5130.7851 &$ 75   \pm 13$& GMOS-S \\
 5130.7873 &$ 81   \pm 13$& GMOS-S \\
 5141.6567 &$ 40   \pm 13$& GMOS-S \\
 5141.6638 &$ 32   \pm 13$& GMOS-S \\
 5141.6709 &$ 37   \pm 13$& GMOS-S \\
 5141.6780 &$ 30   \pm 13$& GMOS-S \\
\hline
\end{tabular}
\end{table}

\subsection{Spectroscopy}
 We obtained 9 spectra of \J\ on four nights with the GMOS spectrograph
on the Gemini-South telescope using the 600 line/mm grating and
a 0.5\,arcsec slit. We only used the data from the CCD
covering the wavelength range 3698\,--\,4619\AA\ for this study. The
resolution of these spectra estimated from a Gaussian fit to an arc line is
approximately 2.5\AA\ and the dispersion is 0.9\AA\ per pixel. We also
obtained 6 spectra with the EFOSC2 spectrograph on the ESO NTT telescope with
grism \#19. These spectra cover the wavelength range 4434\,--\,5109\AA\ at a
dispersion of 0.67\AA\ per pixel and have a resolution of approximately
2.2\AA. Finally, we obtained 2 consecutive spectra of \J\ with the ISIS
spectrograph on the 4.2-m WHT. These spectra cover the wavelength range
3984\,--\,4776\AA\ at a dispersion of 0.22\AA\ per pixel and have a resolution
of 0.6\AA. 

 The flux-calibrated spectrum of \J\ is compared to the spectrum of the A6Vp
star HD148898  in Fig.~\ref{spec}. 
% which has an effective temperature $\Teff \approx
%8400$ \citep{2007MNRAS.374..664C}. 

% For the ISIS and GMOS spectra used the optimal extraction algorithm of
%\citet{1986PASP...98..609H}  to extract the spectra from the images.

\begin{table}
 \caption{Circular orbit fit for \J\,A. The function fitted to the radial
velocities in Table~\ref{rvtable} from $\gamma + K_A\sin[2\pi({\rm HJD}-{\rm T}_0)/P]$}
\label{rvfitTable}
 \begin{tabular}{@{}lrl}
\hline
  Parameter &\multicolumn{1}{l}{Value} &  \\
\hline
  ${\rm T}_0$ & 2455135.295& (fixed) \\
  P [d]  & 0.66783& (fixed) \\
  $\gamma$ [\kms] & 52.6  & $\pm 2.4$ \\
  $K_A$ [\kms]& 38.5  & $\pm 3.6$ \\
  $N$ & 17 \\
  $\chi^2$ & 12.2 \\
\hline
\end{tabular}
\end{table}

\section{Analysis}

 We refer to the larger, cooler component of \J\ as \J\,A and its
companion (the star eclipsed at phase 0)  as \J\,B. 

\subsection{Lightcurve model}
 In addition to the 3 lightcurves we obtained, we also analysed the V-band
photometry for \J\ provided by the  ASAS survey \citep{ 2002AcA....52..397P}. We used only the
503 measurements graded ``A'' for our analysis. We used the lightcurve model
{\sc ebop} \citep{1981psbs.conf..111E,1981AJ.....86..102P} to analyse all four
lightcurves simultaneously  and so derive the following parameters: the radii
of the stars relative to their separation, $R_{\rm A}/a$ and $R_{\rm B}/a$;
the inclination, $i$; the ratio of the surface brightnesses for the stars at
each wavelength, $S_{\rm WASP}$, $S_{\rm ASAS}$, $S_{V}$ and $S_{I_C}$; the
linear limb-darkening coefficients in the V-band and I$_{\rm C}$-band ,
$x_{V}$ and $x_{I_C}$, respectively; the time (HJD UTC) of mid-primary eclipse,
${\rm T}_0$; the orbital period, P; and the normalization of the 4
lightcurves. For the WASP data we use the average of $x_{V}$ and $x_{I_C}$ as
the linear limb darkening parameter.   We assigned the same linear
limb-darkening coefficient to both stars because the limb darkening of
\J\,B  has a negligible effect on the lightcurve. For numerical stability and
to avoid non-physical values for the various parameters, the free parameters
we use in the least-squares fit are: $\log ( S_{\rm WASP})$; $\log(S_{\rm
ASAS})$; $\log(S_{V})$; $\log(S_{I_C})$; $\log (1/b-1)$ (where $b = a\cos
i/[R_{\rm A}+R_{\rm B}]$ is the impact parameter); $R_{\rm A}/a$; $k=R_{\rm
B}/R_{\rm A}$; $(P-0.66783173)\times1000$; ${\rm T}_0 - 2455134.6272$;
$\log(1/x_V-1)$; $\log(1/x_{I_C}-1)$; $\log(q)$ (where $q = M_B/m_A$ is the
mass ratio).  The optimum values of the parameters of interest derived by
least-squares are given in Table~\ref{lcfitTable}. The standard errors on the
parameters are derived used a bootstrap Monte Carlo method. The fits to the
lightcurves can be seen in Fig.~\ref{lcfit}. The joint distributions from the
Monte Carlo simulation for selected pairs of parameters are shown in
Fig.~\ref{plotpar6}. The linear limb-darkening parameters $x_{V}$ and
$x_{I_C}$ are found to be indeterminate from the lightcurve and the other
parameters have only a weak dependence on them. 

\subsection{Effective temperatures of \J\,A and \J\,B}

 We used the observed V$-$K  colour of \J\ combined with the luminosity
ratio and surface brightness ratio in the V-band from the lightcurve solution
to estimate the effective temperatures of \J\,A and \J\,B.  For the V-band
magnitude we used the mean value of V from the ASAS photometry, with the
sample standard deviation as a standard error, $12.28 \pm 0.07$. This standard
error is intended to account for the variation of the flux between eclipses
and any systematic offset from the ASAS V-band and the standard V-band
photometric system. We use the observed value of $K_s$ from Table~\ref{mags}
but as the phase at which this magnitude was measured is not known we assign
it the same standard error as the sample standard deviation of the WASP
photometry. We also convert the K$_s$ magnitude to K using (K$_s$)$_{\rm
2MASS} = {\rm K} - 0.044$ \citep{2005ARA&A..43..293B} to obtain
K=$11.83\pm0.05$.

 The effective temperature of a single star with V$-$K = $0.45$ is
8250\,K \citep{2007hsaa.book.....Z}. We used the spectral energy
distributions from \cite{1993KurCD..13.....K} to calibrate the relation
between surface brightness and effective temperature in the V-band. We then
used the initial value of  ${\rm T}_{\rm eff,A} = 8400$\,K for \J\,A and the
values of $S_V$ from the lightcurve solution to obtain ${\rm T}_{\rm eff,B} =
14050$\,K for \J\,B.  We then use the V$-$K value for a star of this effective
temperature and the luminosity ratio in the V-band from the lightcurve
solution to derive a value of (V$-$K)$_A = 0.51\pm0.09$. Repeating the steps
above then leads to the improved effective temperature estimates of  ${\rm
T}_{\rm eff,A} = 8060\pm330$\,K for \J\,A and ${\rm T}_{\rm eff,B} =
13400\pm1200$\,K for \J\,B. 

 We did not attempt a detailed analysis of our spectra for \J\ because we were
not able to clearly identify any spectral lines from \J\,B. This makes it very
difficult to account for the contamination of the combined spectrum by \J\,B,
particularly since this highly evolved star may have a very peculiar
atmospheric composition, e.g., it may be hydrogen deficient. We were able to
estimate that the projected rotational velocity of \J\,A is $V_A\sin i =
95\pm5$\,\kms\ from the observed widths of various metal lines.   

\subsection{Spectroscopic orbit of \J\,A}
 We measured the radial velocity of  \J\,A by cross-correlation of the spectra
against a high-resolution spectrum of the A6p-type star HD148898
\citep{2003Msngr.114...10B}. We excluded the broad Balmer lines from the
cross-correlation. There is no indication of a second peak in the cross
correlation function due to \J\,B. The radial velocities derived from the
cross correlation function are given in Table~\ref{rvtable}. These radial
velocities have been corrected for the radial velocity of the template star 
(2.5\,\kms; \citealt{1953GCRV..C......0W}). For well-exposed
spectra such as these the dominant sources of error are systematic, e.g.,
motion of the star in the slit and instrument flexure. From experience we find
that using dispersion/5 gives a reasonable estimate of the standard error in
these cases, so this is the value given in Table~\ref{rvtable}.

The parameters of the spectroscopic orbit assuming zero eccentricity in the
least-squares fit are given in Table~\ref{rvfitTable}. 

\subsection{Kinematics}
\begin{figure} 
\includegraphics[width=0.45\textwidth]{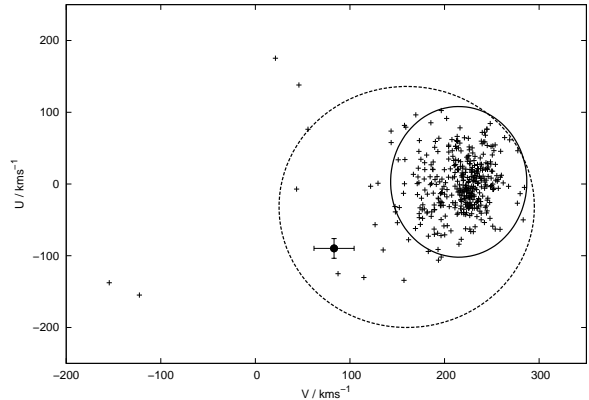} 
\caption{Galactic U-V velocities for \J\ (error bar) compared to the sample of
white dwarfs from \citet{2006A&A...447..173P}. The
contours define the 3-$\sigma$ ellipse for thin disk (full drawn) and thick disk
(dashed) stars as defined by \citeauthor{2006A&A...447..173P}.
\label{uv}  } 
\end{figure} 

 We have calculated the Galactic U-V velocity components of \J\ 
using the method described by \citet{2003A&A...400..877P}.  We used the mean
of the proper motion values given in Table~\ref{mags} and assigned an error of
1.5\,mas to each component. The radial velocity of the system is taken from
Table~\ref{rvfitTable}.  We used the isochrones from
\citet{2000A&A...358..593S} to estimate that a zero-age main-sequence star
with the same effective temperature as \J\,A  has an absolute V magnitude
M$_{\rm V} = 1.4 \pm 0.3$. We calculated the apparent V magnitude of \J\,A
assuming that it contributes 89\,per~cent of the light in the V-band. From the
apparent distance modulus of \J\,A and ignoring the effects of reddening we
estimate the distance to \J\ to be $d=1500\pm 300$\,pc. The resulting U-V
velocity components for \J\  are shown in Fig.~\ref{uv} compared to the region
of this diagram occupied by thin-disk and thick-disk stars. \J\ clearly has
the kinematics of a thick-disk star, which suggests that it is likely to be
old ($\ga 7$\,Gyr), metal poor ($-1 \la [{\rm Fe/H}] \la -0.3$) and have
enhanced $\alpha$-element abundance ([Mg/Fe] $\ga 0.3$).

\begin{figure} 
\includegraphics[width=0.45\textwidth]{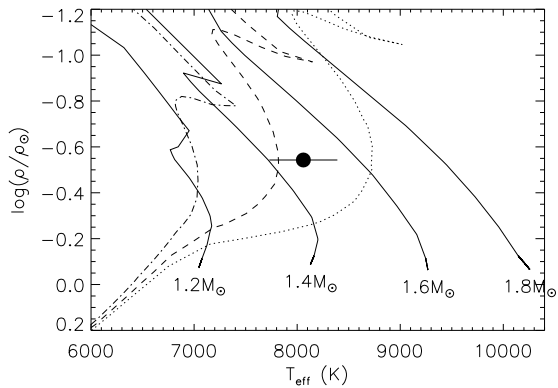} 
\caption{The effective temperature and density of \J\,A compared to the models
of \citet{2000A&AS..141..371G} for Z=0.004. Stellar evolutionary tracks are
shown as thick lines and labelled by mass. Isochrones for $\log({\rm
age/Gyr})= 9.0, 9.2, 9.4$ are plotted with dotted, dashed and dash-dotted lines,
respectively.
\label{trho}  } 
\end{figure} 

\begin{figure} 
\includegraphics[width=0.45\textwidth]{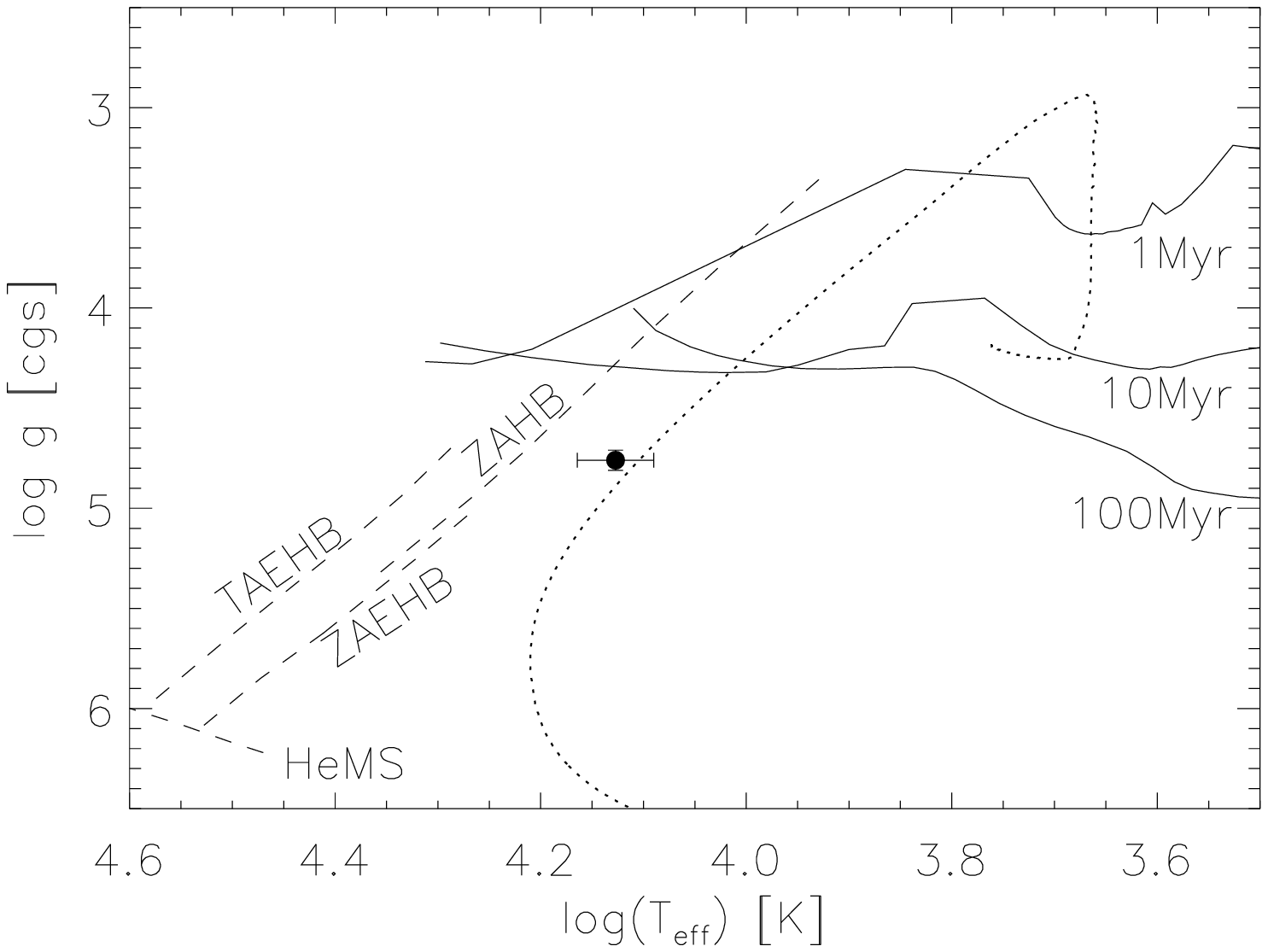} 
\caption{The effective temperature and surface gravity of \J\,B compared to
the models for core hydrogen burning stars (solid lines), core helium burning
stars (dashed lines) and a model for the formation of a 0.195\Msolar\ white
dwarf (dotted line, \citealt{1999A&A...350...89D}). Core hydrogen burning
models from \citet{2000A&A...358..593S} are labelled by age. Core helium
burning models are labelled as follows: ZAEHB/TAEHB = zero-age/terminal-age
extreme horizontal branch \citep{1993ApJ...419..596D}; ZAHB = zero-age
horizontal branch \citep{1987ApJS...65...95S}; HeMS = helium main sequence
\citep{1993ApJ...419..596D}.
\label{tlogg}  } 
\end{figure}

\begin{figure} 
\includegraphics[width=0.45\textwidth]{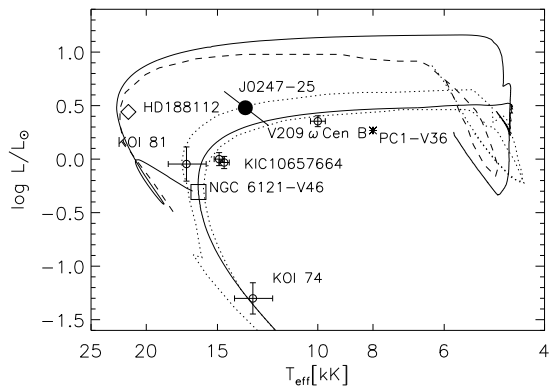} 
\caption{\J\,B in the Hertzsprung-Russell diagram (filled circle). Other
objects discussed in the text are also plotted. Models for the formation of
low mass white dwarfs (with final masses as noted, bottom-to-top)  are also
shown as follows: \citet{1999A&A...350...89D} -- solid lines (0.195\Msolar\ and
0.234\Msolar); \citet{2004ApJ...616.1124N} -- dotted lines (0.205\Msolar\ and
0.215\Msolar); \citet{2010ApJ...715...51V} -- dashed lines (0.21\Msolar).
\label{hrd}  } 
\end{figure}

\subsection{Mass, radius and luminosity of the components}
 Using Kepler's 3$^{\rm rd}$ law we find that the density of \J\,A, $\rho_A$, is
related to the orbital period,  the parameter $R_A/a$ derived from the
lightcurve model and the mass ratio, $q = M_B/M_A$, as follows: 
\begin{equation}
 \rho_A = \frac{3\pi}{G(1+q)(R_A/a)^3P^2}. 
\end{equation}
 The value of $\rho_A$ is not sensitive to the exact value of $q$ provided
this value is small, so we take a value of $q=0.16\pm0.08$  and
compare the values of $\rho_A$ and T$_{\rm eff,A}$ to the stellar models for
main-sequence stars from \cite{2000A&AS..141..371G}. This value of $q$ is
chosen for consistency with the masses derived below. The error on $q$ is
arbitrary but large enough to easily cover all likely values for this
parameter. For the models shown in Fig.~\ref{trho} the  best estimate of the
mass is $1.48\pm0.09$\Msolar\, where the error includes the uncertainties in
T$_{\rm eff,A}$, $\rho_A$ and $[{\rm Fe/H}] = -0.65 \pm 0.35$.

 In Table~\ref{mrl} we give the values for the mass, radius and luminosity for
\J\,A and \J\,B derived from the lightcurve model and mass function assuming
$M_A = 1.48\pm0.09$\Msolar. 

\begin{table}
 \caption{Physical parameters of the  components of \J.}
\label{mrl}
 \begin{tabular}{@{}lrr}
\hline
  Parameter& \multicolumn{1}{l}{\J\,A} &  \multicolumn{1}{l}{\J\,B}   \\
\hline
Mass (\Msolar) & $1.48\pm0.09^{\makebox[0pt][l]{a}}$ & $0.23\pm
0.03^{\makebox[0pt][l]{a}}$ \\
Radius (\Rsolar) & $1.71 \pm 0.04^{\makebox[0pt][l]{a}}$ & $ 0.33 \pm
0.01^{\makebox[0pt][l]{a}}$ \\
$\Teff$ (K) & $8060\pm 330$ & $13400 \pm 1200$\\
$\logg$ [cgs] & $4.13 \pm 0.02^{\makebox[0pt][l]{a}}$ & $4.75 \pm 0.05$ \\
Luminosity (\Lsolar) & $ 11\pm 2^{\makebox[0pt][l]{a}}$   & $ 3 \pm
1^{\makebox[0pt][l]{a}}$\\
\hline
\end{tabular}
\newline
$^{\makebox{a}}$assuming \J\,A is a main-sequence star with $[{\rm Fe/H}] =
-0.65\pm0.35$.
\end{table}

\section{Discussion}

 Our estimate of the surface gravity for \J\,B, $\log g_B$, is independent of
the assumed mass for \J\,A \citep{ 2004MNRAS.355..986S} so in Fig.~\ref{tlogg}
we compare the observed values of T$_{\rm eff, B}$ and  $\log g_B$ to various
models. It is clear that \J\,B cannot be a core hydrogen burning star and is too
cool to be a core helium burning star. In contrast, the model for the
formation of a 0.195\Msolar\ white dwarf from \citet{1999A&A...350...89D} is a
good match to the observed values of T$_{\rm eff, B}$   $\log g_B$ and a
reasonable match to the estimated value of $M_B$. We therefore conclude that
\J\,B is the precursor of a low mass white dwarf. As this white dwarf will be
composed almost entirely of helium, we refer to \J\,B and similar stars as
pre-He-WD stars.

 In Fig.~\ref{hrd} we show the Hertzsprung-Russell diagram (HRD) for various
models of the formation of low-mass white dwarfs and the observed positions
of \J\,B and some related objects. The objects KOI-74 and KOI-81 
discussed earlier appear to form an evolutionary sequence with \J\,B according
to these models, with KOI-81 coming towards the end of the pre-He-WD phase
and KOI-74 being near the start of the white dwarf cooling track. KIC~10657664
is a companion to an A-type star also identified from Kepler photometry
\citep{2011ApJ...728..139C}. The Kepler lightcurve of this binary shows a well
defined total eclipse approximately 2\,per~cent deep, a secondary eclipse, and
variations between the eclipses that were identified as the Doppler beaming
(DB) signal due to the orbital motion of the A-star and the ellipsoidal
variation (ELV) due to its tidal deformation.
The mass ratio for the binary can be estimated from either the DB or ELV
signal, but these are found to be inconsistent with one another unless the
A-type star has a mass much lower than expected (0.7\Msolar\ cf. 2.5\Msolar).
HD\,188112 is a single-lined spectroscopic binary star for which the mass
($0.24^{+0.10}_{-0.07}\Msolar$) can be inferred from its Hipparcos parallax
\citep{2003A&A...411L.477H}. The companion to this star is a compact object
and the orbital of the system 0.606585\,d. The object PC1-V36 is a binary star
with an orbital period of 0.8\,d in the globular cluster 47~Tuc
\citep{2008ApJ...683.1006K}. The companion to this very low mass object
(M=0.056\Msolar) is very faint and may be a neutron star. NGC~6121-V46 is also
a binary star in a globular cluster (M4) with an unseen companion
\citep{2006BaltA..15...61O}.

 The object V209~$\omega$~Cen\,B is one component of an eclipsing binary with
a period of 0.83\,d in the globular cluster $\omega$~Cen
\citep{2007AJ....133.2457K}. This star also has a similar mass to \J\,B
($0.144\pm0.008$\Msolar). The other component in this binary star,
V209~$\omega$~Cen\,A, is too hot to be a main sequence star given its mass
(\Teff = 9370\,K, M=0.945\,\Msolar). This casts doubt on our assumption that
the mass of \J\,A can be estimated from models of normal main-sequence stars.
The same doubt applies to the analysis of KIC~10657664 by
\citeauthor{2011ApJ...728..139C}, who also assumed that the A-type primary
star in that binary is a main-sequence star because a mass of 0.8\Msolar\ for
a star with $\Teff\approx9500$\,K is ``not physically plausible''.
\citeauthor{2007AJ....133.2457K} suggest that V209~$\omega$~Cen\,A may be a
white dwarf that has accumulated sufficient mass to re-ignite shell hydrogen
burning. As far as we can ascertian, this intriguing scenario has not been
studied further.

 The position of \J\,B in Fig.~\ref{tlogg} is not affected by the assumed mass
for \J\,A, so an alternative way to estimate the mass of \J\,A is to assume
that \J\,B has a mass of 0.195\Msolar\ from the model of
\citet{1999A&A...350...89D} that is a good match to the observed values of
T$_{\rm eff, B}$ and $\log g_B$. In this case, the mass function implies a
mass of $M_A = 1.2\pm0.2$\,\Msolar, i.e, similiar to the mass of
V209~$\omega$~Cen\,A.  Alternatively, if we assume that \J\,A\ rotates
synchronously, then the observed value of $V_A\sin i = 95\pm5$\,\kms\ combined
with the parameters in Table~\ref{lcfitTable} can be used to infer the value
of the semi-major axis, $a$, and thus $M_A\approx 0.6$\,\Msolar\ and
$M_B\approx 0.13$\,\Msolar. Finally, we can use the mass ratio estimated from
the lightcurve solution together  with the mass function to estimate  $M_A =
2.8\pm1.0$\,\Msolar. This estimate must be regarded with some caution
because the only feature of the lightcurve that depends strongly on $q$ is the
ellipsoidal variation in brightness between eclipses. It is not known at the
time writing whether the algorithm used to remove systematic noise from the
WASP lightcurves may also reduce the strength of variations on timescales of
$\sim $8\,hours, i.e., the duration of observations on a typical night and the
period of the ellipsoidal variation in \J. Although these various estimates of
$M_A$ are not consistent with one another, this does not affect our
main conclusion that \J\,B is a pre-He-WD.

 We have not shown in Fig.~\ref{hrd} the complex evolutionary paths
followed by low mass white dwarfs during hydrogen shell flashes. Some of these
paths match the observed properties of \J\,B rather well, but the timescale
for the evolution through the relevant part of the HRD is generally extremely
short (decades\,--\,centuries), so this is a rather remote possibilty.

J0247$-$25\,B contributes about 11\% of the flux at visible wavelengths so
better quality spectroscopy should make it possible to directly measure the
radial velocity of \J\,B and so derive precise, model-independent masses for
both stars.  It will also be possible to recover the individual spectra of the
two components of \J\ by comparing the spectra in-eclipse and out-of-eclipse
or using spectral disentangling techniques \citep{2010ASPC..435..207P}.
This will make it possible to test the prediction of
\citeauthor{1999A&A...350...89D} that a 0.195\Msolar\ white dwarf should have
a helium-enhanced atmosphere at this stage of its evolution as a consequence
of the extreme mass loss that this star has suffered. 

\begin{figure} 
\includegraphics[width=0.45\textwidth]{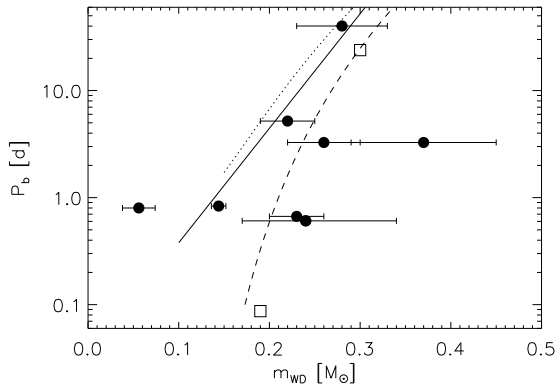} 
\caption{The masses and  periods for low-mass white dwarfs and
pre-He-WDs from Table~\ref{lmwd} compared to selected models as follows:
solid line -- \citet{2004ApJ...616.1124N}; dotted line --
\citet{1995MNRAS.273..731R}; dashed line --  \citet{1999A&A...350..928T}.
Star for which the mass estimates have no quoted error bar are shown with open
symbols.
\label{pmwd}  } 
\end{figure}

 Models for the formation of LMWD suggest that the relationship between
orbital period and white dwarf mass ($P_b - M_{\rm WD}$ relation) should be
almost independent of the details of how the mass is lost from the red giant,
although the relation is expected to show some dependence on metalicity. In
practice, these models seem to under-estimate the mass of LMWD companions to
milli-second pulsars \citep{2005ApJ...632.1060S}. The masses and periods of
various low-mass white dwarf and pre-He-WD binary systems are given in
Table~\ref{lmwd} and are compared to selected models for their formation
through a phase of stable mass transfer in Fig.~\ref{pmwd}.  The relation
between mass and period is very steep at the short period end, so precise and
accurate (model independent) mass measurements are needed to define an
empirical $P_b - M_{\rm WD}$ relation. Spectroscopy of J0247$-$25 with
sufficient signal-to-noise and resolution to allow a spectroscopic orbit for
\J\,B to be measured should yield a precise mass estimate for J0247$-$25\,B.
This would make possible an interesting comparison with V209~$\omega$~Cen\,B
and PC1-V36, which have similar orbital periods to \J\, but that are both
members of metal-poor globular clusters.

\begin{table}
 \caption{Masses and periods for low mass white dwarfs and pre-He-WDs in
binary systems.  } 
 \label{lmwd}
 \begin{tabular}{@{}lrrr}
  \hline
  Name &\multicolumn{1}{l}{Period [d]} & \multicolumn{1}{l}{Mass[\Msolar]} &  Source \\
  \hline
NGC 6121-V46 & 0.087 & $\sim 0.19$ & 1 \\
HD 188112    & 0.607 & $0.24^{+0.10}_{-0.07} $ & 2 \\
\J & 0.668 & $0.23 \pm 0.03 $& 3 \\
PC1-V36      & 0.794 &  $0.056 \pm 0.018$ & 4,5 \\
V209~$\omega$~Cen\,B & 0.834  & $0.144 \pm 0.008$ & 6 \\
KIC 10657664 & 3.274 &  $0.26 \pm 0.04$ & 7 \\
             &        &  $0.37 \pm 0.08$ & 7 \\
KOI-74 & 5.189 & $0.22 \pm 0.03$ &  8 \\
KOI-81 & 23.89 & $\sim 0.3$      &  8 \\
Regulus B    & 40.11  & $0.28\pm0.05$  & 9 \\
\hline
\end{tabular}
\newline
References: 1 -- \citet{2006BaltA..15...61O}; 
 2 -- \citet{2003A&A...411L.477H}; 3 -- this paper; 4 --
\citet{2001ApJ...559.1060A}; 5 -- \citet{2007AJ....133.2457K}; 6 --
\citet{2007AJ....133.2457K}; 7 -- \citet{2011ApJ...728..139C}; 8 --
\citet{2010ApJ...715...51V}; 9 -- \citet{2009ApJ...698..666R}
\end{table}

 It seems clear that the formation of \J\ must have involved extensive mass
loss from a red giant star, but the mechanism for the mass loss is not so
clear. The progenitor of \J\,B must have had a mass $\ga 0.8\Msolar$ to evolve
off the main-sequence within the lifetime of the Galaxy, so this star has lost
$\ga 0.5\Msolar$. This suggests that \J\,B has accreted rather a lot of
material or the evolution of this binary system has required  highly
non-conservative mass transfer.  A full exploration of the possible
evolutionary pathways for the formation of \J\ is beyond the scope of this
paper, but we note here that the low projected rotational velocity of \J\,A
may be a very useful constraint in any such study. If the mass of \J\,B is
$\ga 0.2\Msolar$ then \J\,A must be rotating sub-synchronously or have
significantly non-zero obliquity. This would be difficult to explain in any
scenario in which \J\,A has gained a large amount of mass and angular momentum
from the red giant progenitor to \J\,B. Nevertheless, \J\,A appears to be a
young star when compared to the isochrones for metal-poor stars shown in
Fig.\ref{trho} (1-2\,Gyr), certainly much younger than a typical star in the
thick-disk population \citep{2008PhST..133a4031F}. This suggests that \J\,A
has gained enough mass from its companion to become a blue-straggler, i.e., an
anomolously young, massive star when compared to other thick-disk stars. The
synchronisation timescale for a 1.4\Msolar\ star is about 10\,Myr
\citep{2004A&A...424..919C}, comparable to the time since the formation of
\J\,B according to the 0.195\Msolar\ model of
\citeauthor{1999A&A...350...89D}. This  suggests that there has not been
sufficient time for \J\,A to have lost a large amount of rotational angular
momentum through tidal interactions with \J\,B since its formation. It may be
that the formation of \J\,B left  \J\,A far from equilibrium and that the slow
rotation is caused by the subsequent expansion of this star. 

\section{Conclusions}

 The star  1SWASP~J024743.37$-$251549.2  is an eclipsing binary star in which
the precursor to a low mass white dwarf with a mass $\approx
0.25$\Msolar\ is totally eclipsed by a larger, cooler star once every
0.6678\,d. More detailed spectroscopy will be required to measure a precise
masses for the stars. This will enable us to determine the nature of the
larger star and to make detailed tests of models for the formation 
pre-He-WDs and low mass white dwarfs.

\section*{Acknowledgments}
 Funding for WASP comes from consortium universities and from the UK's
Science and Technology Facilities Council.

WASP-South is hosted by the South African Astronomical Observatory and we are
grateful for their ongoing support and assistance. 

This publication makes use of data products from the Two Micron All Sky
Survey, which is a joint project of the University of Massachusetts and the
Infrared Processing and Analysis Center/California Institute of Technology,
funded by the National Aeronautics and Space Administration and the National
Science Foundation.

 Based on observations obtained at the Gemini Observatory, which is operated
by the Association of Universities for Research in Astronomy, Inc., under a
cooperative agreement with the NSF on behalf of the Gemini partnership: the
National Science Foundation (United States), the Science and Technology
Facilities Council (United Kingdom), the National Research Council (Canada),
CONICYT (Chile), the Australian Research Council (Australia), Minist\'{e}rio
da Ci\^{e}ncia e Tecnologia (Brazil) and Ministerio de Ciencia, Tecnolog\'{i}a
e Innovaci\'{o}n Productiva (Argentina) (Program ID GS-2009B-Q-96).

\label{lastpage}
\bibliographystyle{mn2e}  
\bibliography{wasp}

\end{document}